\documentclass[twocolumn,prl,preprintnumbers,amsmath,amssymb,superscriptaddress,unsortedaddress]{revtex4}

\usepackage{graphicx}
\usepackage{dcolumn}
\usepackage{bm}

\begin{document}


\title{Input Impedance and Gain of a Gigahertz Amplifier Using a DC SQUID in a Quarter Wave Resonator\footnote{This paper is a contribution of the U.S. government and is not subject to U.S. copyright.}}

\author{Lafe Spietz}
 \email{lafe@nist.gov}
\author{Kent Irwin}
\author{Jos\'e Aumentado}%
\affiliation{National Institute of Standards and Technology, Boulder, Colorado 80305, USA}

\date{\today}

\begin{abstract}
 Due to their superior noise performance, SQUIDs are an attractive alternative to high electron mobility transistors for constructing ultra-low-noise microwave amplifiers for cryogenic use.  We describe the use of a lumped element SQUID inductively coupled to a quarter wave resonator. The resonator acts as an impedance transformer and also makes it possible for the first time to accurately measure the input impedance and intrinsic microwave characteristics of the SQUID. We present a model for input impedance and gain, compare it to the measured scattering parameters, and describe how to use the model for the systematic design of low-noise microwave amplifiers with a wide range of performance characteristics.
\end{abstract}

\maketitle

	The Superconducting Quantum Interference Device (SQUID) is a very-low-noise and low power-dissipation gain element that has been used for amplification of signals from DC to microwave frequencies.  The challenge in the use of SQUIDs above 100 megahertz is to overcome the stray reactance of the SQUID\cite{clarkebook}, which makes it difficult to effectively couple in microwave power.  This is a very important application of SQUIDs, however, because microwave SQUID amplifiers have been shown to have much better noise performance than the quietest available semiconducting amplifiers.  Cryogenic semiconducting amplifiers are often the limiting factor in the overall performance of microwave quantum measurement experiments, which has created a significant demand for lower noise solutions.
	
	Previous workers in the field have constructed shunted DC SQUID amplifiers by using the stray capacitance to their advantage to create a microstrip resonator out of the input coil\cite{mueck1}.  This approach yields significant gain, and noise near the quantum limit\cite{mueck_quantum}; however it has proven difficult to accurately model in detail, and the gain of such amplifiers diminishes significantly for frequencies above 1 gigahertz\cite{mueck_ghz}.  We follow the approach of shrinking the physical size and stray capacitance of the SQUID to the point where it may be treated as a lumped element component at microwave frequencies\cite{prokopenko1}\cite{tarasov}.  We then use a resonator to measure the input impedance and gain of that SQUID and compare with a simple theory.
 
	Our overall design philosophy is to make the different elements of our amplifier as independent from one another as possible, so each part can be optimized and mass-produced.  By using a lumped-element SQUID with a separate impedance transformer we have deconstructed the problem of microwave SQUID amplifier design into separate problems of SQUID design and transformer design.  
	
\begin{figure}
\includegraphics{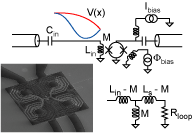}
\caption{Schematic of experiment with transformer model and SEM of SQUID.  A 50 $\Omega$ transmission line is interrupted by a 25-115 fF overlap capacitor and terminated in the input coil of a lumped element SQUID to create a quarter wave resonator.  Our naive circuit model for how power is coupled into the SQUID is shown in the bottom right.  Two DC current supplies at room temperature fix the flux bias and the current bias of the SQUID, and an integrated bias tee at the output allows for simultaneous DC and RF measurement.  The SEM shows the 1.5 turn input coil and the half-turn flux bias coil on the slotted washer of the SQUID.}
\end{figure}

	The experiments described in this paper consist of a series of S-parameter measurements on SQUIDs at the end of a quarter wave resonator built into our fully modular amplifier design mentioned above.  The SQUIDs used in this work are resistively shunted niobium DC SQUIDs in a second-order gradiometer configuration that reduces sensitivity to environmental flux changes, with a slotted washer that minimizes the stray capacitance from the input coil to the SQUIDs\cite{squidref1}\cite{squidref2}\cite{squidref3}(see Fig. 1).  The mutual inductance from the input coil to the SQUID was 40 pH, the SQUID geometric self inductance was 18 pH, and the input coil inductance was calculated to be approximately 600 pH.  In addition, the SQUIDs have flux bias coils that allow for DC flux biasing without having DC access to the main input coil.  The shunt resistors were 2.3 $\Omega$ each, and the junctions had critical currents of 60 $\mu A$ each.  In typical operation, these SQUIDs dissipate approximately 10 nW of power.  

	While construction of wide bandwidth SQUID amplifiers is our final goal, we used narrow-band quarter-wave resonators both to characterize the input impedance of our SQUIDs at various microwave frequencies and to impedance match to 50 $\Omega$ in order to demonstrate microwave power gain.  These resonators consisted of overlap capacitors coupled to lengths of 50 $\Omega$ coplanar waveguide transmission line terminated in the input coil of the SQUID(see Fig. 1).  The overlap capacitors used were 25 fF, 47 fF, and 115 fF, and the resonators were 16 and 25 mm in length.  By having a small capacitor coupled to a 50 $\Omega$ line at one end and an inductor coupled with a small mutual inductance to the low impedance SQUID loop at the other end, the quarter-wave resonator is able to match a very low impedance to a very high impedance by having a current antinode at one end and a voltage antinode at the other end.  A resonator in this configuration has resonances at all odd multiples of the quarter wave frequency, with appropriate shifts in frequency for the reactive elements at the ends.   
	
	This circuit also allows for the characterization of the input impedance of the SQUID at the various resonant frequencies above 1 GHz (SQUID input impedance at lower frequencies has been studied elsewhere \cite{darinZin}).  Looking at $S_{11}$, the reflection coefficient from the input of the device, we observe that almost all microwave power is reflected off resonance, but that there are deep resonances that absorb power.  The Q's of these resonances are determined primarily by the input coupling capacitor, the shifts of the resonances relative to a quarter-wave resonator with no SQUID at the end are determined by the imaginary component of the input impedance, and the depth of the resonance is determined by the real component.  Thus, by measuring $S_{11}$ as a function of current and flux bias, we may determine the real and imaginary components of the input impedance for various resonant frequencies and operating points of the SQUID.  This information is critical both to understanding gain as discussed below in this work, and to designing next-generation lumped-element SQUID amplifiers with targeted operating parameters.  By having two different fundamental frequencies with multiple higher harmonics, we were able to extract the input impedance over a whole range of five discrete frequency values from 1 GHz up to 5.5 GHz.  By varying the capacitance value, we were able to compare power matching to a range of different effective source impedances, giving a higher level of certainty in the final calculated SQUID impedance.

\begin{figure}
\includegraphics{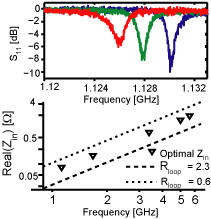}
\caption{Plot of the real component of the input impedance of the SQUID as a function of frequency and return loss of the input resonator for various current bias points.  The dotted and dashed lines represent calculated values from Eq. 1 for maximal and minimal observed values of observed impedance, respectively, and the triangular markers represent observed impedances corresponding to approximately -7 dB power-coupling, which is typical of both optimal gain biasing and of asymptotic behavior at high current biases.  The return loss plots show the raw data from which input impedance is extracted for different bias points of the SQUID.}
\end{figure}

	We now discuss the input impedance model of our DC SQUIDs at microwave frequencies.  Our approach is to construct a simplified model, with the intent of being able to predict input impedance well enough to design amplifiers at targeted frequencies in the gigahertz range.  To this end, we use the simplified circuit shown in the inset of Fig. 1 and make the approximations that $\omega L_{SQ}$, where $L_{SQ}$, the self-inductance of the SQUID loop, is small compared to $R_{loop}$, and that $\frac{M^2\omega^2}{R^2}\frac{L_{SQ}}{L_{in}}<<1.$  These approximations lead to the following approximate formula for input impedance:

\begin{equation}
Z_{in} \approx i\omega L_{in} + \frac{\omega^2M^2}{R_{loop}}.
\end{equation}

	The values of the real component of the input impedance as extracted from the depths of measured $S_{11}$ resonances are shown in Figure 2.  It is clear from examining this plot that impedance matching becomes an \textit{easier} problem at higher frequencies than at lower frequencies, due to the smaller mismatch from 50 $\Omega$.  The extracted loop resistances are lower than four times $R_{dyn}$, the dynamic resistance, as presented in reference \cite{mcdonald_amp_theory}, but are not unreasonable in comparison to the resistances in the SQUID. The expected factor of 4 originates from calculating the series resistance of two parallel resistors.  The input inductances measured from the frequency shifts of the resonances do not all agree on a single value, but agree with a range of inductances from approximately 400 to 600 pH.  Based on this inductance and the real component of the input impedance, we can make a naive estimate of maximum bandwidth based on the ratio of inductance to real input impedance\cite{bode}\cite{fano}\cite{pozar}, implying that it should be possible to get 1 GHz to 5 GHz of bandwidth at higher frequencies with appropriate multipole impedance transformers. 

	In order to understand the gain in this amplifier, we trace the path of incoming microwave power incident on the input of the device on match.  To calculate the gain with imperfect match, we multiply by the power-coupling factors at the input and output, $1 - |S_{11}|^2$ and $1 - |S_{22}|^2$.  To compute the gain we must find how much flux through the SQUID loop the incoming power generates by finding the current through the input coil.  This is found from the real component of the input impedance by using the fact that the power delivered to $Z_{in}$ is $I^2Re[Z_{in}]$, where I is the RMS current driven through the input coil.  From the current through the SQUID input coil, the flux through the SQUID is calculated by multiplying by the input mutual inductance M.  This flux signal constitutes the actual input signal to the SQUID as a gain element, and the intrinsic gain of the SQUID is characterized by the parameter $\partial V/\partial\phi$, the flux to voltage conversion, which has units of hertz.  In the SQUIDs discussed in this paper, $\partial V/\partial\phi$ was generally in the range from 200 to 300 $\mu V/\Phi_0$, (100 to 150 GHz).  Finally, to calculate the gain when the input is matched and the output is out of match, we multiply by $1 - |S_{22}|^2$, the measured output coupling.  This can be calculated approximately from the impedance mismatch from the output impedance of the DC SQUID, which is about 1.5-2 $\Omega$, leading to a loss in gain due to impedance mismatch of approximately 10 dB. 

\begin{figure}
\includegraphics{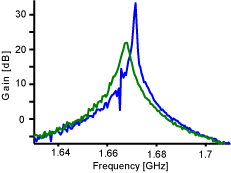}
\caption{Typical gain curves at 1.67 GHz.  Gain curves vary widely in size and shape, and we are working on a more detailed model of gain, which will be presented elsewhere.}
\end{figure}

	Putting all this together, we can finally express our formula for the power gain of the full amplifier as follows:
\begin{equation}
G = \frac{M^2}{Re[Z_{in}]R_{out}}\left(\frac{\partial V}{\partial\phi}\right)^2(1 - |S_{11}|^2)(1 - |S_{22}|^2).
\end{equation}
	This formula can be simplified further by assuming that the real component of the input impedance of the SQUID has the form given in Equation 1, and by assuming a perfect input and output match, which can be achieved with correct microwave design.  This ideal gain is
\begin{equation}
G_{ideal} =  \frac{R_{loop}}{R_{out}}\frac{1}{\omega^2}\left(\frac{\partial V}{\partial\phi}\right)^2.
\end{equation}

	We have measured the gain experimentally, as well as $S_{11}$ and $S_{22}$, $R_{out}$ and $\partial V/\partial\phi$, so that we may compare them with theory.  The theory agrees with the data qualitatively.  The gains measured in the present generation of devices are approximately 26 dB maximum gain at 1.7 GHz and about 19 dB maximum gain at 5 GHz.  The ideal gain formula explains why both our data and other published SQUID amplifier data indicate a decrease in gain with increasing frequency, and why it is so critical to get high values of $\partial V/\partial\phi$ in order to get tolerable gain at the higher frequencies.

	In conclusion, we have presented a general model for the input impedance and gain of a lumped element DC SQUID and described how to achieve an appropriate trade-off between gain and bandwidth for optimal use as a microwave amplifier in a variety of applications.
	
\begin{acknowledgments}
We thank Konrad Lehnert, Michel Devoret, Dan Schmidt, Michael Elsbury, and Rob Schoelkopf for useful discussions.
\end{acknowledgments}

\bibliography{squidrefs}

\begin{thebibliography}{14}
\expandafter\ifx\csname natexlab\endcsname\relax\def\natexlab#1{#1}\fi
\expandafter\ifx\csname bibnamefont\endcsname\relax
  \def\bibnamefont#1{#1}\fi
\expandafter\ifx\csname bibfnamefont\endcsname\relax
  \def\bibfnamefont#1{#1}\fi
\expandafter\ifx\csname citenamefont\endcsname\relax
  \def\citenamefont#1{#1}\fi
\expandafter\ifx\csname url\endcsname\relax
  \def\url#1{\texttt{#1}}\fi
\expandafter\ifx\csname urlprefix\endcsname\relax\def\urlprefix{URL }\fi
\providecommand{\bibinfo}[2]{#2}
\providecommand{\eprint}[2][]{\url{#2}}

\bibitem[{\citenamefont{Clarke and Braginski}(2004)}]{clarkebook}
\bibinfo{author}{\bibfnamefont{J.}~\bibnamefont{Clarke}} \bibnamefont{and}
  \bibinfo{author}{\bibfnamefont{A.~I.} \bibnamefont{Braginski}},
  \emph{\bibinfo{title}{The SQUID Handbook}} (\bibinfo{publisher}{John Wiley
  and Sons}, \bibinfo{year}{2004}).

\bibitem[{\citenamefont{M\"{u}ck et~al.}(1998)\citenamefont{M\"{u}ck,
  Andr\'{e}, Clarke, Gail, and Heiden}}]{mueck1}
\bibinfo{author}{\bibfnamefont{M.}~\bibnamefont{M\"{u}ck}},
  \bibinfo{author}{\bibfnamefont{M.-O.} \bibnamefont{Andr\'{e}}},
  \bibinfo{author}{\bibfnamefont{J.}~\bibnamefont{Clarke}},
  \bibinfo{author}{\bibfnamefont{J.}~\bibnamefont{Gail}}, \bibnamefont{and}
  \bibinfo{author}{\bibfnamefont{C.}~\bibnamefont{Heiden}},
  \bibinfo{journal}{Applied Physics Letters} \textbf{\bibinfo{volume}{72}},
  \bibinfo{pages}{2885} (\bibinfo{year}{1998}).

\bibitem[{\citenamefont{M\"{u}ck et~al.}(2001)\citenamefont{M\"{u}ck, Kycia,
  and Clarke}}]{mueck_quantum}
\bibinfo{author}{\bibfnamefont{M.}~\bibnamefont{M\"{u}ck}},
  \bibinfo{author}{\bibfnamefont{J.~B.} \bibnamefont{Kycia}}, \bibnamefont{and}
  \bibinfo{author}{\bibfnamefont{J.}~\bibnamefont{Clarke}},
  \bibinfo{journal}{Applied Physics Letters} \textbf{\bibinfo{volume}{78}},
  \bibinfo{pages}{967} (\bibinfo{year}{2001}).

\bibitem[{\citenamefont{M\"{u}ck et~al.}(2003)\citenamefont{M\"{u}ck, Welzel,
  and Clarke}}]{mueck_ghz}
\bibinfo{author}{\bibfnamefont{M.}~\bibnamefont{M\"{u}ck}},
  \bibinfo{author}{\bibfnamefont{C.}~\bibnamefont{Welzel}}, \bibnamefont{and}
  \bibinfo{author}{\bibfnamefont{J.}~\bibnamefont{Clarke}},
  \bibinfo{journal}{Applied Physics Letters} \textbf{\bibinfo{volume}{82}},
  \bibinfo{pages}{3266} (\bibinfo{year}{2003}).

\bibitem[{\citenamefont{Tarasov et~al.}(Jun 1995)\citenamefont{Tarasov,
  Prokopenko, Koshelets, Lapitskaya, and Filippenko}}]{prokopenko1}
\bibinfo{author}{\bibfnamefont{M.}~\bibnamefont{Tarasov}},
  \bibinfo{author}{\bibfnamefont{G.}~\bibnamefont{Prokopenko}},
  \bibinfo{author}{\bibfnamefont{V.}~\bibnamefont{Koshelets}},
  \bibinfo{author}{\bibfnamefont{I.}~\bibnamefont{Lapitskaya}},
  \bibnamefont{and}
  \bibinfo{author}{\bibfnamefont{L.}~\bibnamefont{Filippenko}},
  \bibinfo{journal}{Applied Superconductivity, IEEE Transactions on}
  \textbf{\bibinfo{volume}{5}}, \bibinfo{pages}{3226} (\bibinfo{year}{Jun
  1995}), ISSN \bibinfo{issn}{1051-8223}.

\bibitem[{\citenamefont{Tarasov and Ivanov}(Jun 1996)}]{tarasov}
\bibinfo{author}{\bibfnamefont{M.}~\bibnamefont{Tarasov}} \bibnamefont{and}
  \bibinfo{author}{\bibfnamefont{Z.}~\bibnamefont{Ivanov}},
  \bibinfo{journal}{Applied Superconductivity, IEEE Transactions on}
  \textbf{\bibinfo{volume}{6}}, \bibinfo{pages}{81} (\bibinfo{year}{Jun 1996}),
  ISSN \bibinfo{issn}{1051-8223}.

\bibitem[{\citenamefont{Zimmerman and Frederick}(1971)}]{squidref1}
\bibinfo{author}{\bibfnamefont{J.~E.} \bibnamefont{Zimmerman}}
  \bibnamefont{and} \bibinfo{author}{\bibfnamefont{N.~V.}
  \bibnamefont{Frederick}}, \bibinfo{journal}{Applied Physics Letters}
  \textbf{\bibinfo{volume}{19}}, \bibinfo{pages}{16} (\bibinfo{year}{1971}).

\bibitem[{\citenamefont{Drung et~al.}(2007)\citenamefont{Drung, Assmann, Beyer,
  Kirste, Peters, Ruede, and Schurig}}]{squidref2}
\bibinfo{author}{\bibfnamefont{D.}~\bibnamefont{Drung}},
  \bibinfo{author}{\bibfnamefont{C.}~\bibnamefont{Assmann}},
  \bibinfo{author}{\bibfnamefont{J.}~\bibnamefont{Beyer}},
  \bibinfo{author}{\bibfnamefont{A.}~\bibnamefont{Kirste}},
  \bibinfo{author}{\bibfnamefont{M.}~\bibnamefont{Peters}},
  \bibinfo{author}{\bibfnamefont{F.}~\bibnamefont{Ruede}}, \bibnamefont{and}
  \bibinfo{author}{\bibfnamefont{T.}~\bibnamefont{Schurig}},
  \bibinfo{journal}{IEEE Trans. Appl. Super.} \textbf{\bibinfo{volume}{17}},
  \bibinfo{pages}{699} (\bibinfo{year}{2007}).

\bibitem[{\citenamefont{Dantsker et~al.}(1997)\citenamefont{Dantsker, Tanaka,
  and Clarke}}]{squidref3}
\bibinfo{author}{\bibfnamefont{E.}~\bibnamefont{Dantsker}},
  \bibinfo{author}{\bibfnamefont{S.}~\bibnamefont{Tanaka}}, \bibnamefont{and}
  \bibinfo{author}{\bibfnamefont{J.}~\bibnamefont{Clarke}},
  \bibinfo{journal}{Applied Physics Letters} \textbf{\bibinfo{volume}{70}},
  \bibinfo{pages}{2037} (\bibinfo{year}{1997}).

\bibitem[{\citenamefont{Kinion and Clarke}(2008)}]{darinZin}
\bibinfo{author}{\bibfnamefont{D.}~\bibnamefont{Kinion}} \bibnamefont{and}
  \bibinfo{author}{\bibfnamefont{J.}~\bibnamefont{Clarke}},
  \bibinfo{journal}{Applied Physics Letters} \textbf{\bibinfo{volume}{92}}
  (\bibinfo{year}{2008}).

\bibitem[{\citenamefont{McDonald}(1984)}]{mcdonald_amp_theory}
\bibinfo{author}{\bibfnamefont{D.~G.} \bibnamefont{McDonald}},
  \bibinfo{journal}{Applied Physics Letters} \textbf{\bibinfo{volume}{44}},
  \bibinfo{pages}{556} (\bibinfo{year}{1984}).

\bibitem[{\citenamefont{Bode}(1945)}]{bode}
\bibinfo{author}{\bibfnamefont{H.~W.} \bibnamefont{Bode}},
  \emph{\bibinfo{title}{Network Analysis and Feedback Amplifier Design}}
  (\bibinfo{publisher}{Van Nostrand}, \bibinfo{year}{1945}).

\bibitem[{\citenamefont{Fano}(1950)}]{fano}
\bibinfo{author}{\bibfnamefont{R.~M.} \bibnamefont{Fano}},
  \bibinfo{journal}{Journal of the Franklin Institute}
  \textbf{\bibinfo{volume}{249}}, \bibinfo{pages}{57} (\bibinfo{year}{1950}).

\bibitem[{\citenamefont{Pozar}(2005)}]{pozar}
\bibinfo{author}{\bibfnamefont{D.~M.} \bibnamefont{Pozar}},
  \emph{\bibinfo{title}{Microwave Engineering}} (\bibinfo{publisher}{John Wiley
  and Sons}, \bibinfo{year}{2005}).

\end{thebibliography}

\end{document}